\documentstyle[12pt]{article} 
\begin{document} 
\baselineskip 0.7cm 
\def\be{\begin{equation}} 
\def\en{\end{equation}} 
\def\bear{\begin{eqnarray}} 
\def\enar{\end{eqnarray}} 
\title{The Origin of the Entropy in the 
Universe} 
\author{\it Alexander Feinstein\thanks{e-mail: 
wtpfexxa@lg.ehu.es}\\ 
        \it and\\ 
        \it Miguel Angel P\'erez Sebasti\'an\thanks{e-mail: 
             wtbpesem@lg.ehu.es} \\ 
        \it Dpto. F\'\i sica Te\'orica, \\ 
        \it Universidad del Pa\'\i s Vasco, \\ 
        \it Bilbao, Spain} 
\date{\today} 
\maketitle 
\begin{abstract} 
We discuss the entropy generation in quantum tunneling of a relativistic
particle under the influence of a time varying force   
with the help of squeezing formalism. It is shown that if one associates 
classical coarse grained entropy to the phase space volume, 
there is an inevitable entropy increase due to the changes 
in position and momentum variances. The entropy change can be quantified 
by a simple expression $\Delta S=\ln\cosh 2r$, where $r$ is the squeeze 
parameter measuring the ``height" and ``width" of the potential barrier. We  
suggest that the universe could have acquired 
its initial entropy in a quantum squeeze from ``nothing" and briefly 
discuss the implications of our proposal. 
\end{abstract} 
 
One of the major problems of theoretical cosmology is to explain the  
present day large scale isotropy and homogeneity of the universe. The  
possible explanation of this puzzle must have an intimate relation to  
the initial conditions at the beginning of expansion. In a beautiful  
paper \cite{penrose}, Penrose suggested a notion of a geometrical  
gravitational entropy related to the Weyl tensor in order to explain  
the ``low" matter entropy of the present day universe \cite{penrose2}. In  
this picture the universe expands from a highly regular initial state  
characterized by the vanishing Weyl tensor. The irregularities then  
develop with time due to gravitational clamping and the entropy  
grows.  
 
To shed more light on the notion of the gravitational entropy related  
to cosmological expansion two different approaches were adopted.  
Davies \cite{davies} used the analogy of the cosmological horizon to  
that one of the black holes. In the context of the black hole physics  
a well-established Bekenstein-Hawking entropy relation exists connecting  
the area of the event horizon of the black hole to its entropy. The  
status of the cosmological horizon is different, however, and being  
an observer dependent quantity its significance as the measure of  
gravitational entropy is less clear.  
 
In a different approach, a  
number of authors \cite{hu-pavon}-\cite{nesteruk-ottewill}  
have sought the way to relate the entropy of the gravitational field  
to the capacity of the later to create matter particles. 
 
In this paper we take rather a different view on the origin of the  
gravitational entropy in the universe. We suggest that the universe  
could have acquired its  
initial entropy in the process of quantum tunneling from ``nothing"  
\cite{vil1}-\cite{vil3} thus relating the initial entropy of the  
universe to its quantum origin.  
 
In quantum cosmology one describes the whole spacetime by a wave  
function rather than by a classical spacetime \cite{hart-haw, haw}. The birth  
of the universe is described as a quantum tunneling process where the small  
universe nucleates out of ``nothing". This process  
will be referred to, later on, as to quantum squeezing. While  
introducing squeeze formalism does not lead to a new description of  
quantum tunneling, it makes the relation of tunneling with entropy  
generation more transparent.  
 
The entropy generation in quantum tunneling process was first  
discussed by Casher and Englert \cite{casher-englert} who have  
concluded that even in the non-relativ\-istic quantum mechanics, the  
tunneling can generate entropy, identified with the Legendre transformation 
of the Euclidean action in the forbidden region, and yield thermal states with 
negative  temperature. The authors appealed in their paper to the apparent  
unitarity break-down and information loss in the tunneling of a clock.  The 
unitarity break-down is only apparent, as authors pointed out, and  is based 
on the fact that in tunneling one does not account for the  ``backward" waves 
generated on the right side of the barrier by tracing  them out. There is an 
information loss about the system, but it is rather an artifact of  
applying the semiclassical approximation. One could restore the  
unitarity in principle by trying to keep the ``backward" waves in  
mind.  
 
Does this mean that the entropy generation in a quantum tunneling  
problem is not real? Not at all. In physics there are many examples  
where one concentrates on some particular degrees of freedom  
observing the growth of entropy. For instance, if one focuses on  
molecular degrees of freedom, entropy does grow in a chemical  
reaction (the vessel is heating up). The unitarity still strictly  
holds but is rather confined to same hidden degrees of  
freedom. 
 
Our first task in this paper will be to show how the entropy generation  
in quantum tunneling is rather a manifestation of a more general 
intrinsic property of Quantum Mechanics: the classical entropy growth due to a  
quantum squeeze. 
 
In an ordinary Quantum Mechanics (see \cite{merzbacher}, for  
example), when two Hermitian operators do  
not commute $[A,B]=iC$, the product of their uncertainties satisfies  
the following relation 
\be 
\Delta A\Delta B \ge \vert\frac{1}{2}\langle C\rangle\vert. 
\label{incer} 
\en 
The inequality (\ref{incer}) is the manifestation of the essential  
nature of Quantum Mechanics, the indeterministic character of  
the theory. The basic principle of Quantum Mechanics states that it is  
impossible to simultaneously measure, with arbitrary precision, two  
complementary variables since the product of their uncertainties is  
bounded from below.  
 
In fact, the equality in the expression (\ref{incer}) only holds when  
the operators $A$ and $B$ are proportional, so that the states with  
minimal uncertainty satisfy the following equation 
\be 
(B-\langle B\rangle)\,\psi = \frac{i}{2}\,\frac{\langle C\rangle} 
{(\Delta A)^2} \, (A-\langle A\rangle)\psi.  
\en 
If, for simplicity, the operator $A$ is identified with the position  
and $B$ with the linear momentum operators, this equation reads 
\be 
\left(\frac{\hbar}{i}\,\frac{d}{dx} - \langle p_x\rangle\right)\psi =  
\frac{i\hbar}{2(\Delta x)^2}\, (x-\langle x\rangle)\psi. 
\label{mineq} 
\en 
The solution to the equation (\ref{mineq}) gives the following normalized  
wave function 
\be 
\psi(x) = [2\pi (\Delta x)^2]^{-1/4} \,  
\exp\left[-\frac{(x-\langle x\rangle)^2}{4(\Delta x)^2} +  
      \frac{i\langle p_x\rangle x}{\hbar}\right], 
\label{packet} 
\en 
for which the minimum uncertainty relation holds: 
\be 
\Delta x\Delta p = \frac{1}{2}\hbar.  
\en  
 
The states of the form (\ref{packet}) minimizing the uncertainty  
relation are called coherent states in Quantum Optics (see Schumaker  
\cite{schumaker} and the references there in).  
 
Consider now a typical Gaussian wave function of the form given by 
(\ref{packet}) 
\begin{equation} 
\psi(x) = A\,e^{-\frac{1}{2}\gamma x^2} 
\label{gau}. 
\end{equation} 
This type of wave packets arise naturally in systems governed by
quadratic Hamiltonians. 
It follows \cite{schumaker}, that the complex parameter $\gamma$ is  
of most importance, since the form of the wavefunction (\ref{gau})  
suggests that it is an eigenfunction of the linear combination $\tilde x +  
\frac{i}{\gamma}\tilde p$, so that the parameter $\gamma$ is related  
to the variances of position and linear momentum operators. 
 
The linear combination $\tilde x + \frac{i}{\gamma}\tilde p$ can be 
written alternatively with the help of the usual creation and annihilation 
operators as 
\be 
\tilde x + \frac{i}{\gamma}\tilde p= \frac{1}{\sqrt{2}} 
\left[(1+\frac{1}{\gamma})\, a\, + \, (1-\frac{1}{\gamma})\, 
a^\dagger\right]. 
\label{combi} 
\en 
Thus, if one is interested in the dynamical evolution of the 
wavefunction (\ref{gau}) under some quantum fluctuation one may as well 
study the behavior of the operators $a$ and  
$a^\dagger$ instead.  
 
Suppose now, the evolution of the state is governed by a quadratic Hamiltonian, 
one then may introduce \cite{schumaker} a unitary operator, the  
so-called single-mode squeeze operator 
\be 
S_1(r,\varphi) = \exp\left[ \frac{1}{2}r(e^{-2i\varphi}\,a^2- 
e^{2i\varphi}\,a^{\dagger 2})\right],  
\label{squop} 
\en 
where $0\le r<\infty$ and $-\frac{\pi}{2}\le\varphi <\frac{\pi}{2}$, are 
parameters depending on the specific form of the disturbance the initial 
state is subjected to.  
 
A generic evolution of a typical coherent Gaussian state involves displacement  
and rotation together with squeeze. For the purposes of our discussion,  
however, since neither rotation nor displacement produce changes in  
position and momentum uncertainties, 
we concentrate here on the action of squeeze operator alone. 
 
The squeeze operator acts on the annihilation operator in the following 
way  
\be 
S_1(r,\varphi)\,a\,S_1^\dagger(r,\varphi) = a \cosh r + 
a^\dagger\,e^{2i\varphi}\,\sinh r, 
\en 
and this implies that the uncertainties in $a$ and $a^\dagger$  
($\tilde x$ and $\tilde p$) are changed  
under the squeeze, displacing the mean position 
or linear momentum. If one starts with a coherent state described by 
the minimal uncertainty relation $\Delta x\Delta p= 1/2$, $(\hbar=1)$ the  
state will evolve under the squeeze to one with \cite{schumaker} 
\begin{eqnarray} 
\langle(\Delta\tilde x)^2\rangle &=& \frac{1}{2}  
       (\cosh 2r\,-\,\cos 2\varphi\,\sinh 2r) \label{incx}\\ 
\langle(\Delta\tilde p)^2\rangle &=& \frac{1}{2} 
       (\cosh 2r\,+\,\cos 2\varphi\,\sinh 2r) \label{incp}\\
\langle\Delta\tilde x\Delta\tilde p\rangle &=& - \frac{1}{2} 
       \sinh 2r\sin 2\varphi. \label{incxp}
\end{eqnarray} 
 
The parameters $r$ and $\varphi$ of the single-mode squeeze operator are 
related to the parameter $\gamma$ of the Gaussian wave function (\ref{gau}) 
in  the following way \cite{schumaker} 
\begin{equation} 
\gamma = \frac{\cosh r \,+\, e^{2i\varphi}\, \sinh r}{\cosh r \,-\,  
e^{2i\varphi}\, \sinh r}. 
\label{gam} 
\end{equation} 
 
Many useful properties of the single-mode squeeze operator may be  
readily found from the properties of the transformation matrix  
$C_{r,\varphi}$ associated to it 
\begin{equation} 
C_{r,\varphi} = \left( \begin{array}{cc} 
                     \cosh r & e^{2i\varphi}\sinh r \\ 
                     e^{-2i\varphi}\sinh r & \cosh r  
                       \end{array} 
                \right), 
\label{ctu} 
\end{equation}  
such that 
\be 
S_1(r,\varphi) \left( \begin{array}{c} 
                         a \\ 
                         a^\dagger 
                      \end{array} \right) S_1^\dagger(r,\varphi) \equiv 
\left ( \begin{array}{c} 
          \alpha(r,\varphi) \\ 
          \alpha^\dagger(r,\varphi) 
        \end{array}  \right) =  
C_{r,\varphi} \left( \begin{array}{c} 
                         a \\ 
                         a^\dagger 
                      \end{array} \right).  
\en 
 
Actually the transformation matrices $C_{r,\varphi}$ can be figured out 
from the requirement that the unitary transformation on $\tilde x, 
\tilde p$ or $a$, preserve their commutation relations. 
 
Suppose now one starts with a coherent state and then the state evolves. 
We assume that the quantum mechanical wave function at large times can 
be accurately described by classical physics, though instead of 
classical trajectory, the system would be rather represented by a 
classical probability distribution \cite{hillery, guth}. Typically, if 
the evolution is governed by a quadratic Hamiltonian, this
will be a Gaussian probability distribution
\be
{\cal P}(x,p) = A \exp\{-\frac{1}{2} [\alpha x^2 + \beta p^2 +2\gamma
xp]\}, 
\en
where $A$ is fixed by normalizing the probability to unity and $\alpha,\,\beta$ 
and $\gamma$ 
are related to the second noise moments of the position and linear 
momentum operators (the coefficient $\gamma$ expresses the correlations between 
the position and the momentum). 

For the classical probability distribution one may define entropy as:
\be
S= - \int {\cal P}(x,p) \, \ln {\cal P}(x,p) \, dx\,dp.
\en
Evaluating the integral and discarding a constant contribution one gets
\be
S = - \frac{1}{2} \ln (\alpha\, \beta - \gamma^2) ,  
\en
or in terms of the second noise moments the expression for the entropy
becomes 
\be
S = \frac{1}{2} \ln (\langle(\Delta x)^2\rangle \langle(\Delta
p)^2\rangle - \langle\Delta x\Delta p\rangle^2_{sym}). 
\en

Since we have started with the coherent state, the initial product of 
uncertainties is given by $\Delta x_0\Delta p_0=1/2$. 
After the state has evolved, the product of uncertainties can 
be evaluated using the expressions (\ref{incx}) - (\ref{incxp}). We thus
get
\be
S = \frac{1}{2} \ln \left( \frac{1}{4} \left[ \cosh^2 2r - \sinh^2 2r
(\cos^2 2\varphi + \sin^2 2\varphi)\right]\right) = \ln \frac{1}{2}, 
\en
and consequently $\Delta S = 0$. 

From a ``purist" point of view this is a perfectly expected result, no
entropy generation is obtained in a squeeze. After all the evolution is
strictly unitary! Note, however, that one starts with a coherent quantum state and then
expects to correctly describe the system classically at late times. 

In the quantum-to-classical transition the coherence must be lost. To
account for the decoherence the system should be coarse-grained. For our
example we will define the coarse-grained entropy in the spirit of Brandenberger, 
Mukhanov and Prokopec \cite{brandenberger} by averaging the second noise moments over the
whole period of the squeeze angle. The averaged variances are: 
\bear
\langle \overline{(\Delta x)^2}\rangle &=& \frac{1}{2} \cosh 2r \\
\langle \overline{(\Delta p)^2}\rangle &=& \frac{1}{2} \cosh 2r \\
\langle \overline{\Delta x\Delta p}\rangle_{sym} &=& 0, 
\enar
and the ``coarse grained" entropy now grows in the squeeze 
\be
\Delta S = \ln\cosh 2r.\label{entro}
\en
 
Although the expression (\ref{entro}) was derived for the initially  
coherent state, one may show that starting with an arbitrary Gaussian  
state a similar expression for the entropy growth will hold. 
 
The relation (\ref{entro}) is our central result. It shows that 
generically there is an entropy generation associated with the quantum 
evolution of a given coherent state. The increase in entropy is related 
with the increase in the product of uncertainties in position and 
momentum and is telling one that a certain amount of the information  
about the state has been irreversibly lost in a quantum-to-classical
transition. 
Quantitatively this increase can be simply expressed in terms of the  
squeeze parameter $r$ as given by the relation (\ref{entro}).  
 
The expression (\ref{entro}) is applicable in principle both for small and large squeeze 
limits. For the large squeeze limit $r\gg 1 $ we get  
\be 
\Delta S = 2r - \ln 2 ,
\en 
which is identical to the expression given by Gasperini and Giovannini 
\cite{gasperini-giovannini} for the entropy generation associated to the 
particle production re-formulated in squeeze formalism. Also, in the 
large squeeze limit it is similar to the expression given by Rosu and 
Reyes \cite{rosu}. Their expression for the entropy growth associated with 
squeeze  and inferred from the information entropy, also known as the 
Shannon-Wehrl  entropy reads 
\be 
S = 1+\frac{1}{2}\ln\sinh^2r. 
\en 
However, it is easy to see that this expression and those given in \cite{brandenberger} 
as well as in \cite{gasperini-giovannini} give negative entropy growth  
for small $r$ which is rather undesirable. Although in the examples 
\cite{brandenberger, gasperini-giovannini} the small $r$ limit is physically meaningless, 
this is not so in the example considered in \cite{rosu}.  
 
We now consider a relativistic massless particle under the action of a time-dependent 
force $dp(t)/dt$. One may show \cite{popov, hu2} that the particle may be 
described by the following wave function  
\be
\psi(x,t) = \varphi(t) \, e^{i p(t) x}, 
\en
where $\varphi(t)$ satisfies the parametric oscillator type equation 
\be
\frac{d^2 \varphi(t)}{dt^2} + p^2(t)\,\varphi(t)=0.
\label{dina}
\en

We assume that the force switches off at $t\to-\infty\, (+\infty)$, and
that the corresponding asymptotic values of the momentum are given by $p_-\, (p_+)$. 

It is well known that one may express the general solution of the equation (\ref{dina}) as a 
combination of two linearly independent basic solutions
\be
\varphi(t) = a\, \chi(t) \,+\, b^\dagger \,\chi^\ast(t), 
\en
where $a$ and $b^\dagger$ are two arbitrary complex constants. 

On the other hand, the general solution $\varphi(t)$ may also be expressed yet in a 
different base
\be
\varphi(t) = c\, \xi(t) \,+\, d^\dagger \,\xi^\ast(t), 
\en
where $c$ and $d^\dagger$ are different complex constants which are
related to the constants $a$ and $b^\dagger$ by the following transformation 
\cite{gri-sid}
\bear
a &=& e^{-i\theta}\,\cosh r\, c - e^{-i(\theta-2\varphi)}\,\sinh r\, d^\dagger 
\label{tr1}\\
b^\dagger &=& -e^{i(\theta-2\varphi)}\,\sinh r\, c+e^{i\theta}\,\cosh r\, d^\dagger .
\label{tr2}
\enar

If the force vanishes asymptotically at $t\to\pm\infty$, the basic
solutions $\chi(t)$ and $\xi(t)$ may be chosen as 
\bear
\chi(t)&\to&\frac{1}{\sqrt{2p_-}}\, e^{-ip_-t},\quad t\to -\infty\\
\xi(t)&\to&\frac{1}{\sqrt{2p_+}}\, e^{-ip_+t},\quad t\to\infty.
\enar

We assume now, as in \cite{casher-englert}, that there are no reflected waves in 
the future region,
$d^\dagger d=0$, this is to say that the particle tunnels in
time, then the probability of such a tunneling $T^2$, is given by the
ratio \cite{merzbacher}
\be
T^2 = \frac{c^\dagger c}{a^\dagger a}.
\en
and one may further express it in terms of the parameter $r$
appearing in the transformations (\ref{tr1}) and (\ref{tr2})
\be
T^2 = \frac{1}{\cosh^2r}. \label{prob}
\en

Therefore, the parameter $r$ is simply related to the tunneling 
probability as discussed by Grishchuk and Sidorov \cite{gri-sid}. 
Yet, there is more to it. To shed more light 
on the physical interpretation of the parameter $r$ we find it necessary 
to consider further quantization of
the theory, i.e. to treat now the wave function $\psi(x,t)$ as a field
operator. 

The equation (\ref{dina}), then, becomes an operator-valued equation and may be
obtained from the following Hamiltonian 
\be
{\cal H} = \hat\pi^\ast \hat\pi + F(t) (\hat\pi^\ast \hat\varphi + 
\hat\varphi \hat\pi^\ast + \hat\pi \hat\varphi^\ast + \hat\varphi^\ast \hat\pi) + 
G(t) \hat\varphi^\ast \hat\varphi ,
\label{hamil}
\en
where $F(t)$ and $G(t)$ are functions determined  by the form of the external
force and $\hat\pi$ is the canonical momentum of the field
$\hat\varphi$. 

The {\it quadratic Hamiltonian} (\ref{hamil}) determines the time evolution of the 
quantum state. In Schr\"odinger picture, this evolution is given by the
following equation
\be
i\hbar\frac{d}{dt}\,\vert\phi\rangle = {\cal H}\, \vert\phi\rangle,
\en
with Gaussian functions been its solutions. 

On transforming the equation (\ref{dina}) into an operator-valued equation, 
the {\bf c}-numbers $a,\, b^\dagger,\, c$ and $d^\dagger$ become operators 
themselves at the same time satisfying an
operator-valued analogue of the transformations (\ref{tr1}) and (\ref{tr2}). 

These operator equations may be obtained, on the other hand, by considering a unitary
evolution in time of the field under the following operator
\be
U(t) = R(\theta)\, S(r,\varphi)
\en
where the operators $R(\theta)$ and $S(r,\varphi)$ are \cite{gri-sid, hu-matz}
\bear
R(\theta) &=& \exp\{-i\theta(t)\, (c^\dagger c + d^\dagger d)\} \\
S(r,\varphi) &=& \exp\{r(t) (e^{-2i\varphi(t)}\, cd -
e^{2i\varphi(t)}\, c^\dagger d^\dagger)\},
\enar
and the evolution of the $a$ and $b^\dagger$ is determined by
\bear
a &=& R^\dagger S^\dagger\, c\, SR\\
b^\dagger &=& R^\dagger S^\dagger\, d^\dagger\, SR.
\enar
We thus see that the parameter $r$ in the equations (\ref{tr1}) and
(\ref{tr2}) can be physically interpreted as the squeeze parameter of
the second quantized theory and is related to the entropy definition 
(\ref{entro})  on one hand, but it is also associated to the tunneling probability 
in the first quantization by equation (\ref{prob}) on the other. 

Thus, to calculate the squeeze parameter the first quantization is
sufficient. However, to understand its physical connection to squeeze 
operator, one must
consider the further quantization of the theory. 

Now, the tunneling probability is completely determined by the squeeze
parameter $r$ and is independent of the squeezing angle $\varphi$. In
the classical region at late times, the knowledge of $\varphi$ is
useless, thus one may justify averaging over all possible values of
$\varphi$. 

Consequently, using Eq. (\ref{entro}),  one expects that a typical coherent 
state will be squeezed on passing through a potential barrier with the entropy 
gain of  
\be 
\Delta S = \ln \cosh 2r. %= \ln \cosh (A\cosh(2/T)) 
\label{entropy}  
\en 
For large $r\gg 1$, one readily gets 
\be 
\Delta S= - 2\ln T +\ln 2 , 
\en 
where $T$ is the transmission amplitude.  
 
Casher and Englert \cite{casher-englert} obtain a similar result, the
difference being, however, in the physical origin of the ``tunneling
entropy". While Casher and Englert argue that the entropy has a thermal
character, we hold that the entropy growth is produced due to a
quantum-to-classical transition decoherence. Quantitatively the
information loss may be evaluated by considering the growth of the phase
space volume of the coarse grained system. 

We now turn to cosmology. 
In quantum cosmology the metric is substituted by a purely quantum  
quantity, the wave function of the universe, which in turn contains  
all the information and the answers to the question one could ask  
about the state of the universe. The wave function of the universe  
is a solution to the Wheeler-DeWitt equation, which is an analog of  
the Schr\"odinger equation. While in general it looks impossible to  
deal with this equation, one may reduce the degrees of freedom of  
the wave function just to two (the scale factor of the universe and  
some matter field) defining the solutions of the Wheeler-DeWitt  
equation on the so-called minisuperspace. Even this simplifying  
procedure would not leave one with the unique wave function for the  
universe. One must still supply the Wheeler-DeWitt equation with the  
appropriate boundary conditions. This is one of the main difficulties  
with the approach of quantum cosmology, for the boundary conditions  
must be based on some physical experience or intuition which we rather  
lack. The simplest and probably the most natural boundary condition  
for the wave function of the universe was proposed by Vilenkin  
\cite{vil3}. He suggested to impose the boundary conditions  
directly on the superspace by requiring that the wave function of  
the universe should contain only the outgoing waves on the boundaries  
of the superspace. This is the so-called tunneling boundary condition.  
 
The basic idea of this approach is that the universe is created from  
nothing, and is based on the fact that the probability of quantum  
creation of a closed spacetime is nonzero. The simplest model to  
consider would be the closed model with a cosmological constant $\Lambda$.  
 
The Wheeler-DeWitt equation for the wave function $\Psi(a)$ 
reads \cite{zeldovich} 
\begin{equation} 
\frac{d^2\Psi(a)}{da^2} + \omega^{2}(a) \Psi(a) = 0 \label{wdw} ,
\end{equation} 
where $\omega^2(a) = - a^2 (1-\Lambda a^2)$ and $a$ represents the  
scale factor of the universe.  
 
One can easily see that this equation is the same as the equation  
(\ref{dina}), with scale factor playing the role of time.   
 
In the classically allowed region $a\ge \Lambda^{-1/2}$ the solution  to the  
equation (\ref{wdw}) is  
\be 
\Psi(a) = \frac{B_1}{\sqrt{\omega(a)}}\,  
             \exp\{i\int_{\Lambda^{-1/2}}^{a}\omega(a')da'\} 
     \, + \frac{B_2}{\sqrt{\omega(a)}}\,  
             \exp\{-i\int_{\Lambda^{-1/2}}^{a}\omega(a')da'\}. 
\en 
The condition that at $a\to\infty$ one has only expanding universes leads to  
the following solution \cite{vil3} 
\begin{equation} 
\Psi(a) = \frac{1}{\sqrt{\omega(a)}}\, 
\exp\{i\int_{\Lambda^{-1/2}}^{a}\omega(a')da' - i\pi/4\}, \hbox{ for } 
a>\Lambda^{-1/2},
\end{equation} 
note that this condition is equivalent to the absence of reflected waves to the 
right of the barrier in our previous example. 

The squeeze parameter $r$ for this tunneling may be evaluated as  
\be 
r=\ln2 + \frac{1}{3\Lambda} = \ln 2 + \frac{3}{16 G^2 \rho_v},  
\en 
where we have substituted the vacuum energy density for $\Lambda$.  
 
Thus the entropy gained by the universe in the course of the quantum squeeze  
will be  
\be 
\Delta S= \frac{3}{8G^2\rho_v} + \ln 2. 
\label{final}  
\en 
 
After the tunneling the universe is described by the classical solution to  
the Einstein equations with a positive cosmological constant, the de  
Sitter solution. Note, that the entropy generated in a quantum squeeze  
from ``nothing" agrees with the entropy associated with the de Sitter  
solution obtained previously by various authors. Arguing in a somewhat
different way, G. Horwitz \cite{horwitz} \footnote{We are very grateful
to Professor Horwitz for communicating his results to us.} comes to a
similar idea of interconnecting the tunneling entropy to the very
interesting question of the origin of time in quantum cosmology. 
 
One would certainly like to start with a universe having a rather small  
tunneling entropy. This implies that the initial vacuum energy density  
should be as large as possible. At any rate, one may try to put a lower  
bound for the vacuum energy density by arguing that the universe should  
have had acquired less entropy in the squeeze than the present day  
entropy of the universe. Estimating the present day entropy in the  
universe as $10^{100}$ \cite{penrose2}, one may conclude that $\rho_v$  
should have been larger than $10^{-24} GeV^4$ initially. The value  
$10^{-24} GeV^4$ is the lowest possible value the $\rho_v$ could have ever  
had.  
 
For a typical SUSY theory, $\rho_v$ is of order $10^{12} GeV^4$ which  
would leave one with an appreciable entropy of $10^{64}$. This is rather a  
large entropy to start with, however, larger values of $\rho_v$ would  
give a lower value.  
 
One could try to put the lower limit on the initial radius of the  
universe by using the Bekenstein entropy bound for a closed  
system \cite{bekenstein}, however, it is clear that unless we know more  
about fundamental interactions it is difficult to make much practical  
use, in the context of quantum cosmology, of the entropy value.  
 
Nevertheless, our main point here was to show that quantum systems which  
behave almost classically at large times, may be characterized by an  
entropy increase. The entropy acquired has quite a universal character.\\ 
Either it is a particle creation, universe nucleation, or any other different  
form of a quantum squeeze, it is accompanied by the entropy generation due  
to the increase in uncertainty or in the volume of the phase space  
necessary to describe the system in the classical regime. The existence  
of such a classical regime is a must for our conclusions to be meaningful  
because the classical entropy notion introduced here does not apply for  
the systems which can not be described quasi-classically. For those  
systems, one should probably use different entropy  
measures \cite{deutsch, partovi}. Yet, since many physical systems of  
interest do behave classically or almost classically, the entropy  
generation process as described here, could shed more light on the physics  
of these systems.  
 
\vspace*{0.8cm} 
\noindent{\Large \bf Acknowledgments} 
\vspace*{0.5cm} 
 
A.F. is grateful to Prof. Jacob Bekenstein for correspondence and  
suggestions. The authors are grateful to Prof. J. Ib\'a\~nez for  
discussions. 
This work is supported by the Spanish Ministry of Education  
Grant  (CICYT) No. PB-$93-0507$. M.A.P.S. is also supported by a Spanish  
Ministry of  Education pre-doctoral fellowship No. AP92 $\;30619396$.

\end{document}